\documentclass[fleqn]{article}
\usepackage{graphicx}

\begin{document}
\title{ Decays of the  $\hat\rho(1^{-+})$ 
Exotic Hybrid and $\eta$-$\eta'$ Mixing}

\author{ 
Ailin Zhang$^{a,b}$\thanks{zhangal@theory1.usask.ca} and
T.G.\ Steele$^{a}$\thanks{Tom.Steele@usask.ca}
\\
{\small $^a$ Department of Physics and Engineering Physics, University of 
Saskatchewan}\\ 
{\small Saskatoon, SK, S7N 5E2, Canada}\\
{\small $^b$ Institute of Theoretical Physics, P. O. Box 2735, Beijing,
100080,
P. R. China}}

\date {}
\maketitle

\begin{abstract}
QCD sum-rules are used to calculate 
the  $\hat\rho(1^{-+})\to\pi\eta,~ \pi\eta^{\prime}$ decay widths of the exotic hybrid  
in two different  $\eta-\eta^{\prime}$ mixing schemes.
 In the conventional flavour octet-singlet mixing scheme, the decay 
widths are both found to be small, while  in the recently-proposed  quark 
mixing scheme, the decay width 
$\Gamma_{\hat\rho\to\eta\pi}\approx 250\,{\rm MeV}$ is large 
compared with the decay width
$\Gamma_{\hat\rho\to\eta^\prime\pi}\approx 20\,{\rm MeV}$. 
These results provide some insight into $\eta$-$\eta'$ mixing 
and  hybrid decay features. \\
\vskip 0.5 true cmPACS~ Indices: 14.40.Aq, 14.40.Cs, 11.40.Ha, 12.38.Lg
\end{abstract}


\section{Introduction}
In recent years, progress has been made in the search for exotic mesons 
 such as glueballs, hybrids and multi-quark 
states. Investigation of these states will not only
 extend our understanding of hadronic physics beyond the standard quark model,
but will also provide insight into non-perturbative aspects of QCD.

Independent observations of an
exotic isovector $J^{PC}=1^{-+}$ state are
of particular interest because of its 
special quantum numbers and
decay modes identifying it as a hybrid meson. 
The $\hat\rho(1405)$ with mass $1392^{+25}_{-22}$ MeV, width $333\pm50$ MeV
decaying into $\eta\pi$
was observed by both E852 and Crystal Barrel in very different production
processes  \cite{e852,crys}. The $\hat\rho(1600)$ with
mass $1593\pm 8$ MeV, width $168\pm 20$ MeV, decaying into $\rho\pi$, was also
observed by E852 \cite{e852}. Though the underlying structure of this
state is not yet understood, a new important decay mode  
$\hat\rho \to\eta^{\prime}\pi$ has been recently reported, 
where the exotic resonance has a mass $1597\pm 10$ MeV and a width
$340 \pm 40\pm 50$ MeV  \cite{prl2001}.

Spectra and decay characteristics 
of hybrids
have been studied with a variety of theoretical methods such as the MIT bag model, flux tube
model, potential model, quark-gluon constituent model, QCD sum rules and
lattice gauge theory, but the exploration is neither complete
nor definitive \cite{method}. The exotic $1^{-+}$ hybrid is generally predicted to be the 
lowest lying hybrid state, but predictions of its decay rate 
vary widely in these analyses motivating 
further investigation.

Theoretical studies of 
hybrid decays using three point function  sum-rules  
 have identified 
$\hat\rho(1^{-+})\to \rho\pi$  as the dominant decay mode compared to 
$\hat\rho(1^{-+})\to\pi\eta,~\eta^\prime\pi$  \cite{decay,latorre}.  However,  
the complications associated with $\eta-\eta^{\prime}$ mixing (which can potentially 
include a  gluonic component) is a possible source 
of the  discrepancy between these theoretical investigations 
and  experimental observations which find comparable widths in the $\eta\pi$ and $\rho\pi$ channels.

The mixing of 
$\eta$ and $\eta^{\prime}$ 
 has been the subject of many investigations \cite{eta}. In the  
traditional mixing scheme, the $\eta,~\eta^\prime$ states are regarded as 
superpositions of the flavor octet and the flavor singlet with a single mixing angle which can be determined from physical  processes.
 However, recent investigations  show 
that this picture may not be correct \cite{leutwyler} 
and a new orthogonal quark-content basis 
has been  proposed \cite{feldmann}. 
Decay constants and the mixing angle have been  extracted in this new scheme. 
Further insight into $\eta$-$\eta^{\prime}$ mixing  
can thus be gained from calculation 
of $1^{-+}$ 
hybrid decays to $\eta\pi,~\eta^\prime\pi$ in the two schemes, and comparing with the existing experiments in these channels.   
 
Issues of $\eta-\eta^\prime$ mixing are reviewed in Section 2, including the  interpolating currents needed in the sum-rule method. 
The QCD sum-rules for the decays 
 $\hat\rho \to \pi\eta',\pi\eta$ are developed in Section 3, and
 their analysis  
is given in Section 4. The final Section is reserved for conclusions.

\section{$\eta,\eta^{\prime}$ mixing}
Conventionally, $\eta$, $\eta^{\prime}$ states are
considered as superpositions of an $SU(3)$ flavor octet $\eta_8$ and a
flavor singlet $\eta_0$ with one mixing angle $\theta$ as
\begin{equation}
|\eta\rangle=|\eta_8\rangle\cos\theta-|\eta_0\rangle\sin\theta,~~
|\eta^{\prime}\rangle=|\eta_8\rangle\sin\theta +|\eta_0\rangle\cos\theta.
\end{equation} 
The decay constants are defined by
\begin{equation}
\label{coupling}
\langle 0 | J_{5\mu}^i | P (p)\rangle = i \,
f_P^i \, p_\mu  \qquad (i=8,0; \quad P=\eta, \eta')\, ,
\end{equation}
where $ J_{5\mu}^8$ denotes the $SU(3)_F$ octet and $ J_{5\mu}^0$ the
$SU(3)_F$ singlet axial-vector current.
It is typically assumed that the decay constants follow the
pattern of state mixing:
\begin{eqnarray}\label{parameter}
&& f_{\eta\phantom{'}}^8 = f_8 \, \cos\theta \ , \qquad
   f_{\eta\phantom{'}}^0 = - f_0 \, \sin\theta \ , \cr
&&  f_{\eta'}^8 = f_8 \, \sin\theta \ , \qquad
   f_{\eta'}^0 = \phantom{-} f_0 \, \cos\theta  ~,
\label{oldmix}
\end{eqnarray}
with the decay constants $f_8$, $f_0$ are defined by couplings of $\eta_8$ and
$\eta_0$ to the divergence of the relevant axial vector currents:
\begin{eqnarray}
\partial_{\mu}j^8_{5\mu}&=&{2\over\sqrt{6}}(m_u\bar ui\gamma_5u+m_d\bar
di\gamma_5d-2m_s\bar si\gamma_5s),\\\nonumber
\partial_{\mu}j^0_{5\mu}&=&{2\over\sqrt{3}}(m_u\bar ui\gamma_5u+m_d\bar
di\gamma_5d+m_s\bar si\gamma_5s)+{1\over\sqrt{3}}{3\alpha_s\over
4\pi}G^A_{\mu\nu}\tilde{G}^A_{\mu\nu},
\end{eqnarray}
where $G^A_{\mu\nu}$ is the gluonic field strength tensor and its 
dual is $\tilde{G}^A_{\mu\nu}=\frac{1}{2}\epsilon_{\mu\nu\rho\sigma}{G}^A_{\rho\sigma}$.
Through analyses of physical processes involving 
$\eta,~\eta^\prime$, the mixing angle is respectively found to be 
$\theta\simeq -10^{\circ}$ or $\theta\simeq -23^{\circ}$ according to the 
quadratic and the linear Gell-Mann-Okubo mass formula.

However, recent investigations indicate that the three
parameters $f_8,~f_0,~\theta$ representation (\ref{parameter}) of 
the decay constants defined in (\ref{coupling}) do not seem to follow the same mixing pattern 
as the physical states  
\cite{leutwyler}.
Motivated by the different influences on the flavor structure of the
breaking of $SU(3)_F$ and the state mixing of vector and tensor mesons, a
new $\eta-\eta^{\prime}$ mixing scheme is proposed \cite{feldmann}.
In this scheme, the two orthogonal basis states $\eta_q$ and $\eta_s$ are
related to the physical states by the transformation
\begin{equation}
 \left (\matrix{\eta \cr \eta'}\right )\,=\, U(\phi)\;
                      \left (\matrix{\eta_q \cr \eta_s}\right ) ,
\label{qsb}
\end{equation}
where $U$ is a unitary matrix defined by
\begin{equation}
    U(\phi)\,=\,\left(\matrix{\cos{\phi} & -\sin{\phi} \cr
                                \sin{\phi} &
\phantom{-}\cos{\phi}} \right) ,
\label{uni}
\end{equation}
and ideal mixing corresponds to the case $\phi=0$.

In the same way, the decay constants are defined as
\begin{eqnarray}
\langle 0|j^i_{5\mu}|P(p)\rangle=if^i_P\, p_\mu\quad
(i=q,s;\quad P=\eta,\eta^{\prime}),
\end{eqnarray}
where $j^i_{5\mu}$ denotes the axial vector currents with quark content
$i=q,s$. Their explicit form are
\begin{eqnarray}
j^q_{5\mu}&=&{1\over \sqrt{2}}(\bar u\gamma_{\mu}\gamma_5u+\bar
d\gamma_{\mu}\gamma_5d)\\\nonumber
j^s_{5\mu}&=&\bar s\gamma_{\mu}\gamma_5s.
\end{eqnarray}
with divergences
\begin{eqnarray}
\partial_{\mu}j^q_{5\mu}&=&{1\over\sqrt{2}}[2m_u\bar
ui\gamma_5u+2m_d\bar di\gamma_5d+{\alpha_s\over
2\pi}G\tilde{G}],\\\nonumber
\partial_{\mu}j^s_{5\mu}&=&2m_s\bar
si\gamma_5s+{\alpha_s\over4\pi}G\tilde{G},
\end{eqnarray}
where $G\tilde G\equiv G^A_{\mu\nu} \tilde G^A_{\mu\nu}$.
Through these divergences, the decay constants can also be written as
\begin{equation}
\langle 0 | \partial_\mu j_{5\mu}^i | P \rangle = M_P^2 \,
f_P^i\ .
\end{equation}
Phenomenological analysis in this scheme reveals that
the decay constants follow  the pattern of the state 
mixing 
corresponding to a mixing angle $\phi=39.3^\circ\pm 1^\circ$  \cite{feldmann}.

\section{Sum-Rules for exotic hybrid decays to $\eta\pi,~\eta^\prime\pi$}

Following the sum-rule methods of \cite{decay,latorre}, 
the three point function used for the 
sum rule corresponding to $\hat\rho(1^{-+})\to
\pi\eta', \pi\eta$ is 
\begin{equation}\label{ga}
\Gamma_\mu(p,q)  = 
i\int d^4xd^4y\, e^{i(qx+py)}\langle 0|T
j_{\pi}(x)\partial_{\nu}j^i_{5\nu}(y)O_{\mu}(0)|0\rangle
 =  T_s(p+q)_\mu+T_A(p-q)_\mu
\end{equation}
where
\begin{eqnarray}\label{cur}
j_{\pi}(x)&=&(m_u+m_d):\bar u(x)i\gamma_5d(x):\\\nonumber
O_{\mu}(x)&=&g:\bar u\gamma_\nu T^a dG^a_{\nu\mu}(x):.
\end{eqnarray}
In the flavor octet-singlet mixing scheme, the interpolating 
currents $\partial_\nu j^i_{5\nu}$ for these processes
are $\partial_\nu j^0_{5\nu}$ and $\partial_\nu j^8_{5\nu}$, while in 
the new quark $\eta-\eta^\prime$ mixing scheme,   the appropriate currents are 
 $\partial_\nu 
j^q_{5\nu}$ and $\partial_\nu j^s_{5\nu}$.

At leading order in perturbation theory and to lowest order in the non-strange
quark masses, the quark components of the currents can be ignored in the correlation function compared to the gluonic component. Thus the gluonic component of the currents provides the dominant QCD contribution.   
Since this gluonic contribution to the correlation function is proportional to $(p-q)_\mu$, we focus on the anti-symmetric factor $T_A(p,q)$. 
Using the results of \cite{latorre} 
leads to  the following theoretical expressions for 
$T^i_A$ at the symmetric Euclidean point 
$-p^2=-q^2=-(p+q)^2=Q^2\gg\Lambda^2$
\begin{eqnarray}
T^8_A&=&0,\\
T^0_A&=&{\sqrt{3}\alpha_s(m_u+m_d)\over 3}\left\{
{3\alpha_s\over 2\pi^2}
\langle \bar uu + \bar dd \rangle \ln{Q^2\over \nu^2}
+{1\over 4\pi Q^2}g\langle \bar uGu+\bar dGd\rangle+
{13\over 18Q^4}\langle \bar uu+\bar dd \rangle \langle \alpha_sG^2\rangle
\right\},\\
T^q_A&=&{\sqrt{2}\alpha_s(m_u+m_d)\over 3}\left\{{3\alpha_s\over 2\pi^2}
\langle \bar uu + \bar dd \rangle \ln{Q^2\over \nu^2}
+{1\over 4\pi Q^2}g\langle \bar uGu+\bar dGd\rangle+
{13\over 18Q^4}\langle \bar uu+\bar dd \rangle \langle \alpha_sG^2\rangle
\right\},\\
T^s_A&=&{\alpha_s(m_u+m_d)\over 3}\left\{{3\alpha_s\over 2\pi^2}
\langle \bar uu + \bar dd \rangle \ln{Q^2\over \nu^2}
+{1\over 4\pi Q^2}g\langle \bar uGu+\bar dGd\rangle+
{13\over 18Q^4}\langle \bar uu+\bar dd \rangle \langle \alpha_sG^2\rangle
\right\},
\end{eqnarray}
where $T^i_A$ ($i=0,~8,~q,~s$) are results corresponding to different currents 
in (\ref{ga}).  Note that since $\partial_\nu j^8_{5\nu}$ 
has no gluon anomaly term, the leading-order perturbative contribution to 
$T^8_A$ is suppressed by non-strange quark mass factors
 and can consistently be set to zero within our approximation.
A fascinating aspect of this three-point function is the suppression of perturbative corrections compared 
with nonperturbative effects.

To get the phenomenological representation of the three point function, we
define
\begin{eqnarray}
\langle 0|j_{\pi}(0)|\pi \rangle=f_{\pi}m^2_{\pi},~~
\langle 0|O_\mu(0)|\hat\rho
\rangle=\sqrt{2}f_{\hat\rho}M^3_{\hat\rho}\varepsilon_\mu,
\end{eqnarray}
where $\varepsilon_\mu$ is the polarization vector of the hybrid.
From the discussions in Section 2, the interpolating current 
$\partial_\nu j^i_{5\nu}$ couples to both the $\eta$  and $\eta^\prime$ 
states, indicating that  the saturation with 
one channel in the three point function is insufficient. 
Thus  $\eta$-$\eta'$ mixing effects require 
 saturation of the three point function with two channels: 
$\hat\rho\to\eta\pi$ and $\hat\rho\to\eta^{\prime}\pi$. For simplicity, 
the contributions of single pole terms and high excited
states are omitted in our results, leading to the phenomenological result
\begin{eqnarray}
T^i_A(Q^2)=g_{\hat\rho\eta\pi}{f_\pi m^2_\pi\over 
Q^2+m^2_\pi}{f^i_\eta
M^2_\eta\over
Q^2+M^2_\eta}{\sqrt{2}f_{\hat\rho}M^3_{\hat\rho}\over Q^2+M^2_{\hat\rho}}+
g_{\hat\rho\eta^\prime\pi}{f_\pi m^2_\pi\over Q^2+m^2_\pi}{f^i_{\eta^\prime}
M^2_{\eta^\prime}\over
Q^2+M^2_{\eta^\prime}}{\sqrt{2}f_{\hat\rho}M^3_{\hat\rho}
\over Q^2+M^2_{\hat\rho}},
\end{eqnarray}
where $g_{\hat\rho\eta\pi},~g_{\hat\rho\eta^\prime\pi}$ are the strong 
couplings and $i=0,~8,~q,~s$.

Equating the theoretical side with the phenomenological side and taking a
Borel transformation of $Q^2 T_A$, we obtain
\begin{eqnarray}\label{equ1}
{g^{os}_{\hat\rho\eta\pi}f^8_\eta
f_{\hat\rho}M^3_{\hat\rho}\over
f_\pi}{M^2_\eta\over
M^2_{\hat\rho}-M^2_\eta}\left(e^{-M^2_\eta\tau}-e^{-M^2_{\hat\rho}\tau}\right)
&+&{g^{os}_{\hat\rho\eta^{\prime}\pi}f^8_{\eta^\prime}
f_{\hat\rho}M^3_{\hat\rho}\over
f_\pi}{M^2_{\eta^\prime}\over
M^2_{\hat\rho}-M^2_{\eta^\prime}}\left(e^{-M^2_{\eta^\prime}\tau}-e^{-M^2_{\hat\rho}\tau}\right)=0,
\\
{g^{os}_{\hat\rho\eta\pi}f^0_\eta
f_{\hat\rho}M^3_{\hat\rho}\over
f_\pi}{M^2_\eta\over
M^2_{\hat\rho}-M^2_\eta}\left(e^{-M^2_\eta\tau}-e^{-M^2_{\hat\rho}\tau}\right)
&+&{g^{os}_{\hat\rho\eta^{\prime}\pi}f^0_{\eta^\prime}
f_{\hat\rho}M^3_{\hat\rho}\over
f_\pi}{M^2_{\eta^\prime}\over
M^2_{\hat\rho}-M^2_{\eta^\prime}}\left(e^{-M^2_{\eta^\prime}\tau}-e^{-M^2_{\hat\rho}\tau}\right)
\nonumber\\
&=&{\sqrt{6}\bar\alpha_s(\tau)\over 8\tau^2}\left\{{\bar\alpha_s\over \pi^2}+
{13\over 27}\langle \alpha_sG^2
\rangle\tau^2\right\} ,
\end{eqnarray}
\begin{eqnarray}\label{equ2}
{g^{qm}_{\hat\rho\eta\pi}f^q_\eta
f_{\hat\rho}M^3_{\hat\rho}\over
f_\pi}{M^2_\eta\over
M^2_{\hat\rho}-M^2_\eta}\left(e^{-M^2_\eta\tau}-e^{-M^2_{\hat\rho}\tau}\right)
&+&{g^{qm}_{\hat\rho\eta^{\prime}\pi}f^q_{\eta^\prime}
f_{\hat\rho}M^3_{\hat\rho}\over
f_\pi}{M^2_{\eta^\prime}\over
M^2_{\hat\rho}-M^2_{\eta^\prime}}\left(e^{-M^2_{\eta^\prime}\tau}-e^{-M^2_{\hat\rho}\tau}\right)
\nonumber\\
&=&{\bar\alpha_s(\tau)\over 4\tau^2}\left\{{\bar\alpha_s\over \pi^2}+
{13\over 27}\langle \alpha_sG^2
\rangle\tau^2\right\}
,\\
{g^{qm}_{\hat\rho\eta\pi}f^s_\eta
f_{\hat\rho}M^3_{\hat\rho}\over
f_\pi}{M^2_\eta\over
M^2_{\hat\rho}-M^2_\eta}\left(e^{-M^2_\eta\tau}-e^{-M^2_{\hat\rho}\tau}\right)
&+&{g^{qm}_{\hat\rho\eta^{\prime}\pi}f^s_{\eta^\prime}
f_{\hat\rho}M^3_{\hat\rho}\over
f_\pi}{M^2_{\eta^\prime}\over
M^2_{\hat\rho}-M^2_{\eta^\prime}}\left(e^{-M^2_{\eta^\prime}\tau}-e^{-M^2_{\hat\rho}\tau}\right)
\nonumber\\
&=&{\sqrt{2}\bar\alpha_s(\tau)\over 8\tau^2}\left\{{\bar\alpha_s\over \pi^2}+
{13\over 27}\langle \alpha_sG^2
\rangle\tau^2\right\}
\label{equ3}
\end{eqnarray}
where $\tau$ is the Borel variable and the PCAC relation 
\begin{eqnarray}
m^2_\pi f^2_\pi=-(m_u+m_d)\langle \bar uu+\bar dd \rangle
\end{eqnarray}
has been used. The notation $g^{os}$ is a reminder that the strong couplings are referenced to the octet-singlet mixing scheme, while  $g^{qm}$ indicates the quark mixing scheme.
The above sets of equations can be solved for  the strong couplings which then
lead to  the decay widths through
\begin{eqnarray}\label{decayw}
\Gamma_{\hat\rho \to P\pi}\simeq
|g_{\hat\rho P\pi}|^2{M_{\hat\rho}\over 48\pi}
\left(1-{M^2_P\over M^2_{\hat\rho}}\right)^3.
\end{eqnarray}
Results of this analysis will be presented in the next Section.

\section{Decay widths for $\hat\rho \to \pi\eta',\pi\eta$}
For the analysis of the sum-rules, we use QCD and hybrid meson parameters consistent with 
 \cite{narison}
\begin{eqnarray}
& &\langle \alpha_s 
G^2 \rangle=0.07{\rm  GeV^4}\quad ,\quad \Lambda^{(3)}=350\,{\rm MeV}
\\
& & M_{\hat\rho}=1.6\, {\rm GeV}\quad ,\quad 
f_{\hat\rho}=30\,{\rm MeV} \quad,
\label{rho_hat_params}
\end{eqnarray}
combined with the ref.\ \cite{fazio} values for
the $\eta$ and $\eta'$ decay constants in the two mixing schemes
\begin{eqnarray}
& & f^q_\eta=0.144\,{\rm GeV}~ ,~ 
f^q_{\eta^\prime}=0.125\,{\rm GeV} ~ ,~ f^s_\eta=0.13\,{\rm GeV}, 
f^s_{\eta^\prime}=0.12\,{\rm  GeV}~,\\
& & f^0_{\eta^\prime}=0.178\,{\rm GeV}~,~ 
f^0_\eta=0.011\,{\rm  GeV}~,~ f^8_{\eta^\prime}=0.012\,{\rm GeV}~,~ 
f^8_\eta=0.190\,{\rm GeV}\quad.
\label{coupling_params}
\end{eqnarray}
Other parameters occurring in the sum-rule are completely standard:
$f_\pi=132$ MeV, $M_\eta=0.548$ GeV, $M_{\eta^\prime}=0.958$ GeV.
Via  (\ref{equ1})--(\ref{equ3}),
these parameters completely determine the strong couplings                                  
and hence the decay widths in the two mixing scenarios.
A difficulty associated with this procedure is the large uncertainty associated with 
$f_{\hat\rho}$
($25\,{\rm MeV}<f_{\hat\rho}<50\,{\rm MeV}$) \cite{narison}, and the lack of freedom  to 
consider smaller values of the hybrid mass $M_{\hat\rho}$ that would be consistent with the 
$\hat\rho(1405)$. However, since $f_{\hat\rho}$ is a common scale on the phenomenological side of the sum-rules, ratios of couplings can be obtained independent of $f_{\hat\rho}$ 
with the freedom to vary  $M_{\hat\rho}$.  We will also ignore any 
uncertainties associated with the remaining 
input parameters since the symmetric kinematic point and the simple phenomenological model 
used in the sum-rule introduces uncertainties which are not easily modelled.  
Thus our results should be considered as determining  the characteristic  scales of the decay widths for comparison of the two mixing scenarios.

\begin{figure}[hbt]
\centering
\includegraphics[scale=0.7]{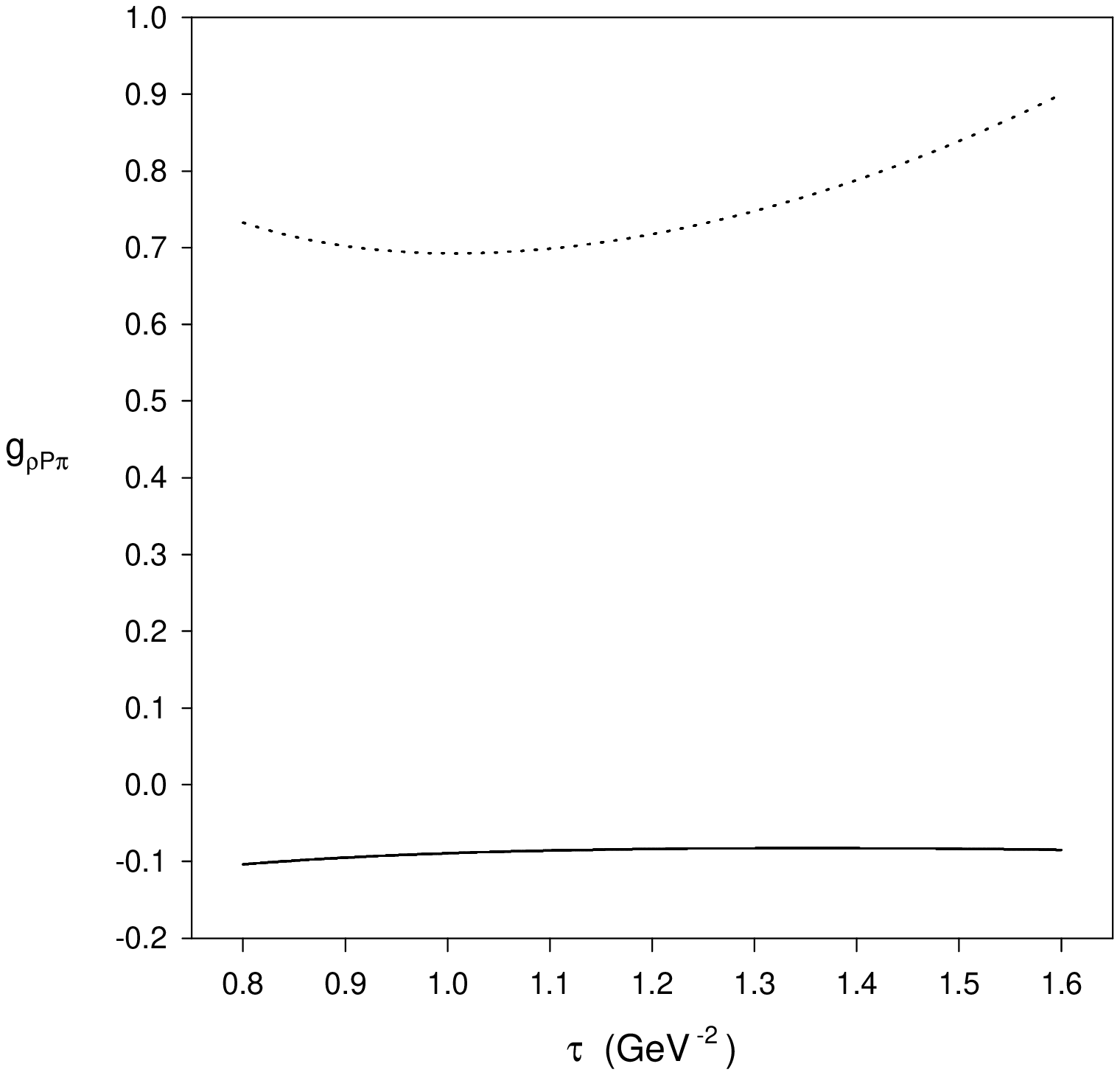}
\caption{
Strong couplings as a function of the Borel parameter $\tau$ in the octet-singlet
$\eta$-$\eta'$ mixing scheme.  The solid curve represents $g^{os}_{\hat\rho\eta\pi}$ and the dashed curve represents $g^{os}_{\hat\rho\eta'\pi}$.  The extremum of the curves provides the sum-rule prediction of the couplings.
}
\label{os_couplings}
\end{figure}

\begin{figure}[hbt]
\centering
\includegraphics[scale=0.7]{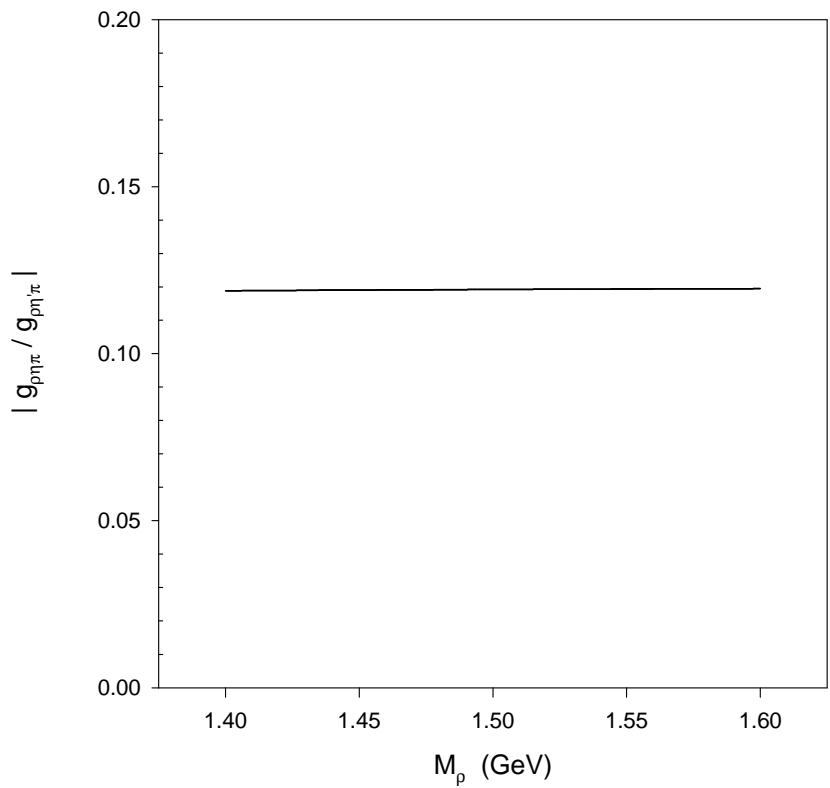}
\caption{
The strong coupling ratio $\left|g^{os}_{\hat\rho\eta\pi}/g^{os}_{\hat\rho\eta'\pi}\right|$ as a function
of the hybrid mass $M_{\hat\rho}$ in the octet-singlet $\eta$-$\eta'$ mixing scheme.
}
\label{os_ratio}
\end{figure}

We begin by considering the octet-singlet mixing scheme.  Figure \ref{os_couplings} displays the
 strong couplings as a function of the Borel scale $\tau$. As necessary in a sum-rule analysis, the
couplings show $\tau$ stability, and there  exists an extremum at a reasonable scale   which then 
defines the sum-rule prediction of the coupling.  For the parameters given in (\ref{rho_hat_params})--(\ref{coupling_params}) we find
extremely small decay widths 
\begin{eqnarray}
\Gamma_{\hat\rho\to\eta^\prime\pi}\simeq 1.4\, {\rm MeV},~~~~~
\Gamma_{\hat\rho\to\eta\pi}\simeq 0.05\, {\rm MeV}.
\end{eqnarray}
These  results are consistent with other sum-rule analyses using the same mixing scheme
which conclude that the widths are no larger than  $10\,{\rm MeV}$  \cite{decay,latorre}.
Figure \ref{os_ratio} displays the ratio of the sum-rule prediction of the strong couplings as a function the hybrid mass.  As noted above, this ratio is independent of the value $f_{\hat\rho}$ from \cite{narison}.  The ratio is remarkably stable at a value of 
\begin{equation}
\left|{g^{os}_{\hat\rho\eta\pi}\over g^{os}_{\hat\rho\eta'\pi}}\right|\simeq 0.12 
\end{equation}
leading to 
\begin{equation}
{\Gamma_{\hat\rho\to\eta\pi}\over 
\Gamma_{\hat\rho\to\eta'\pi}} \simeq 0.04\sim 0.06 
\end{equation}
over the range $1.4\,{\rm GeV}<M_{\hat\rho}<1.6\,{\rm GeV}$.  Thus in the octet-singlet mixing scheme, 
the decay widths for both processes are very small and 
${\Gamma_{\hat\rho\to\eta\pi}\ll \Gamma_{\hat\rho\to\eta'\pi}}$.

\begin{figure}[hbt]
\centering
\includegraphics[scale=0.7]{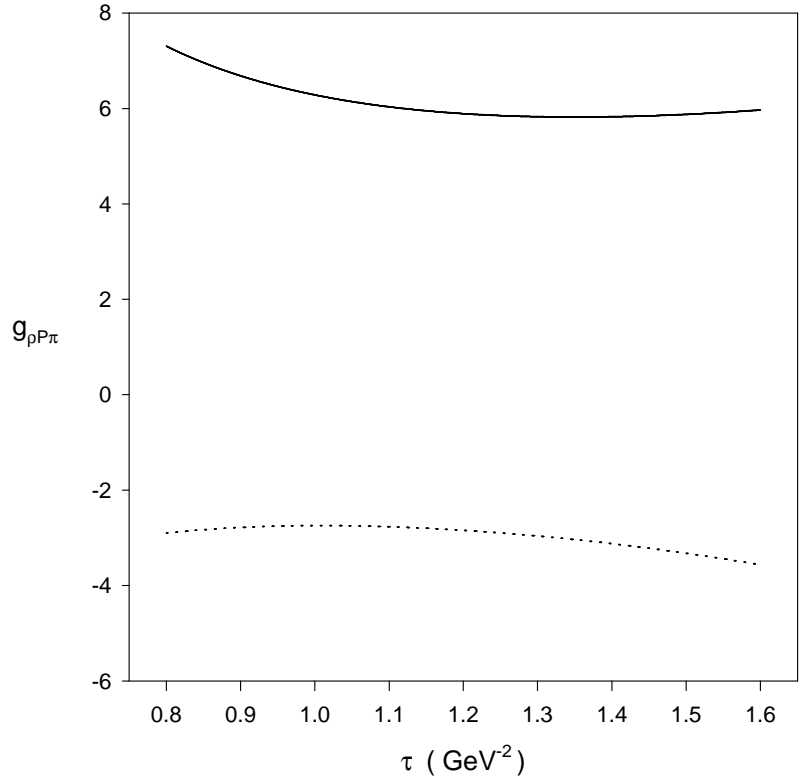}
\caption{
Strong couplings as a function of the Borel parameter $\tau$ in the quark scheme for
$\eta$-$\eta'$ mixing.  The solid curve represents $g^{qm}_{\hat\rho\eta\pi}$ and the dashed curve represents $g^{qm}_{\hat\rho\eta'\pi}$.  The extremum of the curves provides the sum-rule prediction of the couplings.
}
\label{qm_couplings}
\end{figure}

\begin{figure}[hbt]
\centering
\includegraphics[scale=0.7]{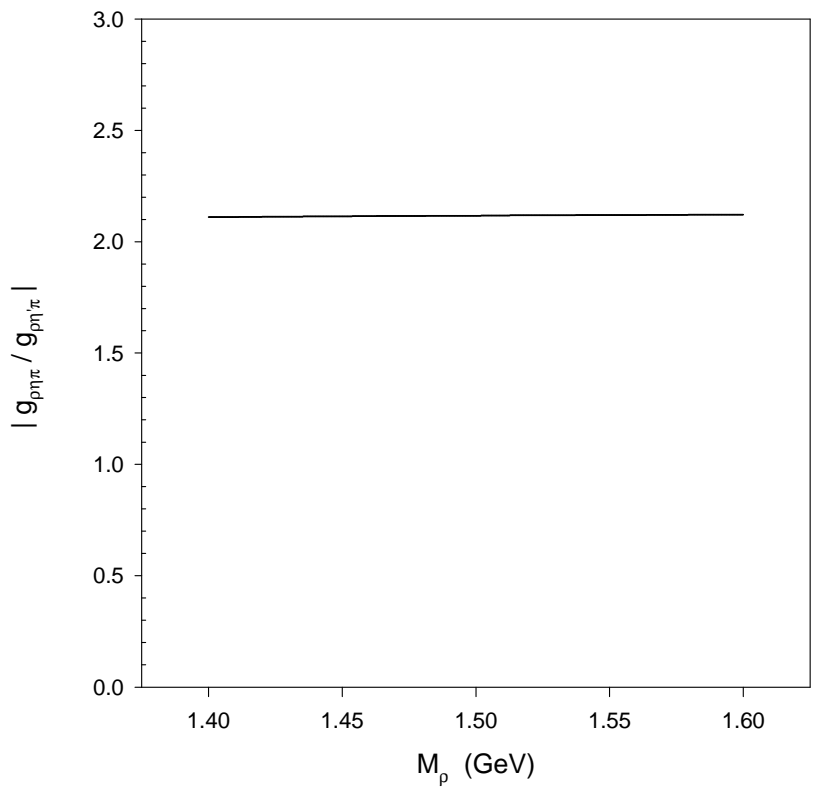}
\caption{
The strong coupling ratio $\left|g^{qm}_{\hat\rho\eta\pi}/g^{qm}_{\hat\rho\eta'\pi}\right|$ as a function
of the hybrid mass $M_{\hat\rho}$ in the quark scheme for  $\eta$-$\eta'$ mixing.
}
\label{qm_ratio}
\end{figure}

The quark mixing scheme leads to substantially different results.  Figure \ref{qm_couplings} displays the
 strong couplings as a function of the Borel scale $\tau$, leading to the
 sum-rule predictions 
\begin{eqnarray}
\Gamma_{\hat\rho\to\eta^\prime\pi}\simeq 21\, {\rm MeV},~~~~~
\Gamma_{\hat\rho\to\eta\pi}\simeq 250\, {\rm MeV}.
\end{eqnarray}
In stark contrast with the octet-singlet mixing scheme, the quark mixing scheme  generates a large width for the 
$\hat\rho\to\eta\pi$ process comparable to that observed experimentally \cite{e852,crys}. 
Figure \ref{qm_ratio}  displays a stable ratio for the sum-rule prediction of the 
strong couplings as a function the hybrid mass
\begin{equation}
\left|{g^{qm}_{\hat\rho\eta\pi}\over g^{qm}_{\hat\rho\eta'\pi}}\right|\simeq 2.1
\end{equation}
leading to 
\begin{equation}
{\Gamma_{\hat\rho\to\eta\pi}\over 
\Gamma_{\hat\rho\to\eta'\pi}} \simeq 10\sim 20 
\end{equation}
over the range $1.4\,{\rm GeV}<M_{\hat\rho}<1.6\,{\rm GeV}$.

Thus we see that the $\eta$-$\eta'$ mixing scheme has a significant impact on the decay widths for the 
processes $\hat \rho\to \eta\pi,~\eta'\pi$. In particular, the quark mixing scheme predicts a large value
for $\Gamma_{{\hat\rho}\to\eta\pi}$  that is comparable to $\Gamma_{{\hat\rho}\to\rho\pi}$ and is consistent with experimental observations.

\section{Conclusions}
In this paper, the sum-rules describing 
 $1^{-+}$ exotic hybrid 
decays  $\hat\rho\to\eta\pi,~\eta^\prime\pi$ have been examined in two different $\eta$-$\eta'$ mixing schemes.

In the singlet-octet scheme, we reproduce the previous sum-rule results \cite{decay,latorre}
which find $\Gamma_{\hat\rho\to\eta\pi}\sim\Gamma_{\hat\rho\to\eta'\pi}\ll\Gamma_{\hat\rho\to\rho\pi}$
which differs from the experimentally observed decay pattern 
$\Gamma_{\hat\rho\to\eta\pi}\sim\Gamma_{\hat\rho\to\eta'\pi}\sim\Gamma_{\hat\rho\to\rho\pi}$ 
\cite{e852,crys,prl2001}.  By contrast, the quark mixing scheme results in  
a significant enhancement of the $\eta\pi$ channel, leading to $\Gamma_{\hat\rho\to\eta\pi}\approx 250\,{\rm MeV}$ comparable to the experimental observations.  However, we find a much smaller width 
in the $\eta'\pi$ channel; $\Gamma_{\hat\rho\to\eta'\pi}\approx 20\,{\rm MeV}$ and in general
$\Gamma_{\hat\rho\to\eta\pi}/\Gamma_{\hat\rho\to\eta'\pi}\sim 15$ contrary to the experimental observation
\cite{prl2001}.

Theoretical refinements extending our analysis beyond the symmetric momentum point and employing a more elaborate phenomenological model could be considered, but it seems unlikely that such refinements would 
generate a sufficiently large value of $\Gamma_{\hat\rho\to\eta'\pi}$ to accommodate the observed value.  
However, there exists evidence for a gluonic component in the $\eta$-$\eta'$ system 
\cite{ball} which would have an impact on our analysis, and could enhance the width of the $\hat\rho\to\eta'\pi$ channel.

Further theoretical work and experimental confirmation of $\Gamma_{\hat\rho\to\eta'\pi}$ are necessary to reach a definitive conclusion, but the quark mixing scheme result   
$\Gamma_{\hat\rho\to\eta\pi}\approx250\,{\rm MeV}$ comparable to the experimental value provides 
support for the quark $\eta$-$\eta'$ mixing scheme rather than the singlet-octet approach.

\section*{Acknowledgments}
Research funding from the Natural Science  \& Engineering Research Council of Canada (NSERC)
is gratefully acknowledged.
Ailin Zhang is partly supported by National Natural Science Foundation of 
China and BEPC National Lab Opening Project.

\clearpage

\clearpage

\clearpage

\clearpage

\end{document}